\documentclass[preprint]{aastex}

\slugcomment{
  Reprint from ApJ, 557, 637--645, 2001.\\
  \copyright\  2001. The American Astronomical Society. All rights reserved.
}

\shorttitle{MIR SED of NGC 1068}
\shortauthors{Tomono et al.}

\begin{document}

\title{Mid Infrared Spectral Energy Distribution of NGC 1068 with
  0\arcsec.1 Spatial Resolution
  \footnote{
    Based on data collected at Subaru Telescope, which is
    operated by the National Astronomical Observatory of Japan.
  }
}
\author{
  Daigo Tomono\altaffilmark{2},
  Yoshiyuki Doi, Tomonori Usuda, and Tetsuo Nishimura
}
\affil{
  Subaru Telescope, National Astronomical Observatory of Japan,
  Hilo, HI 96720
}
\altaffiltext{2}{Current address:
  Max-Planck-Insitut f\"ur extraterrestrische Physik,
  Giessenbachstrasse 1, 85748 Garching, Germany.
}
\email{tomono@mpe.mpg.de}

\begin{abstract}
  The central region of the Seyfert 2 galaxy NGC 1068 is imaged in
  the mid infrared (MIR) using the Mid-Infrared Test Observation System
  on the 8.2 m Subaru Telescope.
  The oversampling pixel scale associated with shift-and-add method shows
  0\arcsec.1 resolution images with a high dynamic range after
  deconvolution. Along with an extended structure at a position angle
  (P.A.) of $-$10\arcdeg\ with higher surface brightness, another
  structure extends wider with
  lower surface brightness at a P.A. of 20\arcdeg. The central peak
  elongates north-south with FWHM of 0\arcsec.3
  $\times$ 0\arcsec.2. Spectral energy distribution (SED)
  of the central peak is fitted to have the silicate absorption
  feature of $\tau_{9.7 \mu m}$ = 0.9 $\pm$ 0.3.
  This is half of the absorption expected from the near-infrared (NIR)
  feature of carbonaceous dust.
  This suggests a temperature gradient of the absorbing
  dust along the line of sight.
  Another possibility, which is not distinguishable here, is the size
  distribution of dust different from our Galaxy.
  Intrinsic luminosity of emission from the central peak is
  3 $\times$ 10$^{37}$ W.
  The SED shows a hint of the poly aromatic hydrocarbon (PAH) emission
  features. Although a high spatial resolution MIR spectrum is required,
  it suggests that the PAH carriers near the active galactic nuclei
  (AGNs) are sheltered from
  the high-energy emission from the AGNs and the AGNs have nuclear starbursts.
  For the NIR disklike structures, no counterparts are detected in the
  MIR.
  The nature of the structures remains unclear.
\end{abstract}

\objectname{NGC 1068}

\section{Introduction}
NGC 1068 is the prototype of Seyfert 2 galaxies
\citep{Antonucci93}. \citet{Antonucci85} detected broad emission lines
in polarized light and confirmed that it has a Seyfert 1 type active
galactic nucleus (AGN) in the center of the galaxy.
The Seyfert 2 AGN is obscured by a dust torus.
There have been a number of observations trying to
detect the plausible torus that has a key to explain the differences
between Seyfert 1 and Seyfert 2 nuclei. For example, adaptive
optics (AO) images by \citet{Marco00} in the L and M bands show
disklike structures in the east and west of the near infrared (NIR;
1--5 $\mu$m) peak of NGC 1068.
However, no counterparts have been observed in the mid infrared (MIR;
8--13 $\mu$m and $\sim$ 20 $\mu$m), which the wavelength
emission from the dust torus would peak. MIR spectra of NGC 1068 and
other AGNs have shown the silicate absorption feature at 9.7 $\mu$m.
This probably
originates in dusty regions relatively close to the central source
\citep{Roche84, Roche91}. When assessing the absorption, care must be
taken because the infrared emitting region might have a spatial
structure and therefore might contaminate the absorption feature
toward the central source.

The advent of the 8 m class
telescopes is expected to generate higher than ever angular resolution
images in the MIR.
\cite{Bock00} observed NGC 1068 using the 10 m Keck II telescope with
a pixel scale of 0\arcsec.138 pixel$^{-1}$.
Using four filters in the MIR, they imaged spatial distribution of
silicate absorption and interpreted that the northern elongation from
the central peak is
reprocessed radiation from the AGN, which is strongly beamed.
The direction of the elongation is
different from that in the MIR images of a wider field of view (FOV) by
\citet{Braatz93} and Marco, Alloin, \& Beuzit (1997).

Using the Mid-Infrared Test Observation System (MIRTOS),
we observed NGC 1068
to clarify the nature of the disklike structures imaged in the NIR
and the difference of elongation in the MIR images.
The 8.2 m Subaru Telescope provides us spatial resolution in the
MIR comparable with that of the NIR AO images.
The shift-and-add (SAA) method delivers diffraction-limited point-spread
function (PSF) in the MIR. The pixel scale of the
MIRTOS is small enough for image reconstruction methods such as
deconvolution.
Six filters are used in the 10 $\mu$m band along with an 18.5 $\mu$m
filter to measure the silicate feature as well as
the polycyclic aromatic hydrocarbon (PAH) band strengths.

The reminder of the paper is organized as follows:
the observation procedures are discussed in \S\ \ref{sec:Observations};
data reduction procedures used to obtain the deconvolved images are
discussed in \S\ \ref{sec:DataReduction};
the resulting data is presented in \S\ \ref{sec:Results};
in \S\ \ref{sec:Discussion}, the MIR radiation from the central peak is
compared with the observations in other wavelengths;
finally, \S\ \ref{sec:Conclusion} contains a summary of our results.
NGC 1068 is at z = 0.0038 \citep{Bottinelli90};
1\arcsec\ corresponds to 72 pc when $H_0$ = 75 km s$^{-1}$ Mpc$^{-1}$.

\section{Observations}
\label{sec:Observations}
Observations of NGC 1068 were made on 1999 December 31,
2000 January 9, and January 18 UT.
We used the MIRTOS mounted at the Cassegrain focus of the 8.2 m Subaru
Telescope.

MIRTOS is constructed for high angular resolution imaging employing
the SAA method to fully extract the diffraction-limited angular resolution of
the telescope. SAA is a method that shifts snapshot
images to cancel the image motion from the atmospheric perturbation before
co-adding the images. MIRTOS combines an MIR camera and an NIR camera
that take simultaneous images in the same field. The MIR camera has an
SBRC (now Raytheon) Si:As array with 320 $\times$ 240 pixels. Its
FOV of 21\arcsec\ $\times$ 16\arcsec\ is provided by a pixel scale of
0\arcsec.067
pixel$^{-1}$. The pixel scale matches $\lambda/3D$ at
$\lambda$ = 8 $\mu$m for $D$ = 8.2 m and produces enough spatial sampling for
deconvolution. The NIR camera
can be used to obtain a reference point source for the SAA method.
Sensitivity in the NIR is higher than in the MIR.
Therefore, the image motion by the atmospheric perturbation can be
measured even with the short integration time that is limited
by the timescale of the image motion.
It has an SBRC InSb array with 256 $\times$ 256 pixels.
The pixel scale of 0\arcsec.028 pixel$^{-1}$ is small enough to detect
the image motion. It provides an FOV of 7\arcsec\ $\times$ 7\arcsec.
As a beam splitter, a long pass interference filter is used to reflect
the NIR light while passing through the MIR light.
To adjust the optical axes of the two cameras to that of the
telescope, the beam splitter and a flat-folding mirror can be moved in the
cryostat. The MIR camera takes images with one of six filters 
from 7.7 through 12.3 $\mu$m or an 18.5 $\mu$m filter
while the NIR camera images in the $J$, $H$, or $K$-bands. Reflective
optics are used in both cameras with a common CdTe dewar entrance window.
Details about MIRTOS are explained in \citet{Tomono00} and
\citet{Tomonoetal00}.

Filters used for this work are listed in Table \ref{tab:Flux}. Because
the target is bright enough, the SAA method was applied using NGC
1068 itself as the
reference in the MIR and in the NIR separately. The integration
time of each frame was made shorter than 0.1 s so that the image
motion does not blur the images.
Secondary mirror chopping was not ready at the time of the observations.
Instead, we performed nodding observations.
Integration for a nodding set was repeated as follows:
\begin{enumerate}
\item Accumulate short exposure frames on the target in the on-source field.
  In a duration of 3.0 s, MIR frames and NIR frames are integrated
  simultaneously. For the MIR, a sequence of 96 frames of 0.031 s integration
  time is acquired consecutively. For the NIR,
  integration time was 0.094 s for each of consecutive 32 frames.
\item Nod the telescope to the offset field.
\item Observe the blank sky in the offset field.
   (The source could be in the offset field.)
\item Nod the telescope back to the on-source field.
\end{enumerate}
Through all the observations, the sky was photometric.
Total time for each nodding cycle was typically 180 s.

For NGC 1068, most of the nodding sets were taken with the offset
field at 30\arcmin\ north of the on-source field.
On January 18, after quick reduction of some of the data that
had been taken, it was confirmed that there is no significant
radiation detected in the MIR around the core of NGC 1068. Therefore,
some nodding sets were taken
with the offset field at 8\arcsec\ east of the on-source field so
that both fields integrated the target in the MIR.
Table \ref{tab:obslog:NGC} shows number of nodding sets taken on each night.
For the observation on January 18, the nodding pairs with 8\arcsec\ offset
are counted as two sets.

The PSF reference sources were observed before
or after the observation of NGC 1068 in each night.
Table \ref{tab:obslog:psfref} shows the number of nodding sets taken
for the PSF references. Typical nodding
offset was 7\arcsec\ east for the MIR and 3\arcsec\ west for the NIR.
All the nodding pairs are
counted twice because the source was integrated in both the on-source field
and the offset field. We did not have a chance to observe a PSF reference
source in 18.5 $\mu$m on January 9. As a substitution,
$\beta$ And, observed on January 18, is used as the PSF reference to
reconstruct the image. The SAA images
of NGC 1068 are essentially the same for the two nights. For some
nights, we also observed other point sources.
To check deconvolution reliability, these data are
processed in the same way as for the NGC 1068 data. Table
\ref{tab:obslog:dectst} shows the number of nodding sets taken for the
deconvolution test sources.

For flux calibration, data with lower time resolution are also used
along with those used for the SAA method.
These data were obtained in the process of pointing the telescope to
the reference sources.
The frame rate was the same for the SAA data, but the frames were
co-added before being transferred to the host computers so that
the transfer time was reduced.
Table \ref{tab:obslog:lumref} shows the number of nodding sets taken
for the flux reference stars.
The brightness of the MIR reference star $\beta$ And is taken from \citet{Cohen95};
the brightness of $\alpha$ CMa is taken from \citet{Tokunaga84},
as well as that of $\alpha$ Ari from \citet{Rieke85}.
$K$-band flux of GJ 105.5 and HD 40335 are taken from \citet{Elias82}.

\section{Data reduction}
\label{sec:DataReduction}
\subsection{Sky Subtraction and Shift-and-Add}
The images are co-added in the following procedure.
First, the foreground emission from the sky and the telescope
is subtracted from every frame.
Frames in the nodding pair are averaged to make the sky frame.
Next, the SAA method is applied on the
images. Images in a nodding pair are co-added by shifting them to
cancel the movement of the images.
The peak
position of the object in each frame is detected after convolving the
frame with the Airy pattern to minimize the noise.
Afterwards, the images are averaged for each night taking the airmass
difference into account.
The 2.1, 11.7, and 12.4 $\mu$m images of NGC 1068 were
taken with a wide range of air-mass for some nights.
We use them to estimate the air-mass dependency of the signal.
Air-mass dependencies at 11.7 and 12.4 $\mu$m were
essentially the same.
Because some of the data are taken at air mass as large as 1.5, the
signals are corrected assuming the estimated air-mass dependency.
We estimate that the brightness conversion has 10\% of uncertainty
in the air-mass correction.
The averaged images for different nights are compared and are
confirm that surface brightness and shape are in a good agreement.
Finally, the images are averaged over all the observation nights.
Figure \ref{fig:R.contour}$a$ shows the resultant SAA image in 11.7 $\mu$m,
for example.

\subsection{Deconvolution}
\label{sec:deconv}
The SAA image in Figure 1$a$ elongates to a position angle (P.A.) of
$-$10\arcdeg.
The elongation is wider in the north than in the south.
The 10 Jy arcsec$^{-2}$ contour spreads to the east by north of the
central peak.
The entire structure elongates from the north by east to the south by
west.
To look into the emission structures, image
reconstruction methods are employed. Two of the image reconstruction
methods, the Lucy-Richardson method \citep{Lucy74} and the maximum
entropy method (MEM; Cornwell \& Evans 1985) are applied using the
IRAF package\footnote{
  IRAF is distributed by the National Optical Astronomy Observatories,
  which are operated by the Association of Universities for Research
  in Astronomy, Inc., under cooperative agreement with the National
  Science Foundation.}.
The Lucy-Richardson method ended with the $\chi^2$ ranging from 1
to 2 after 20 or less iterations.
For the MEM, number of iteration was limited to 110.
Again, resultant images for different nights are compared and
confirm that they have essentially the same morphology.
Moreover, the Lucy-Richardson images and the MEM images show the same
structures,
although the MEM images show more structures with high spatial
frequency.
The high spatial frequency structures are probably artifacts because they
differ from night to night.
Hereafter, the Lucy-Richardson deconvolved images are used.

\begin{figure}
\plotone{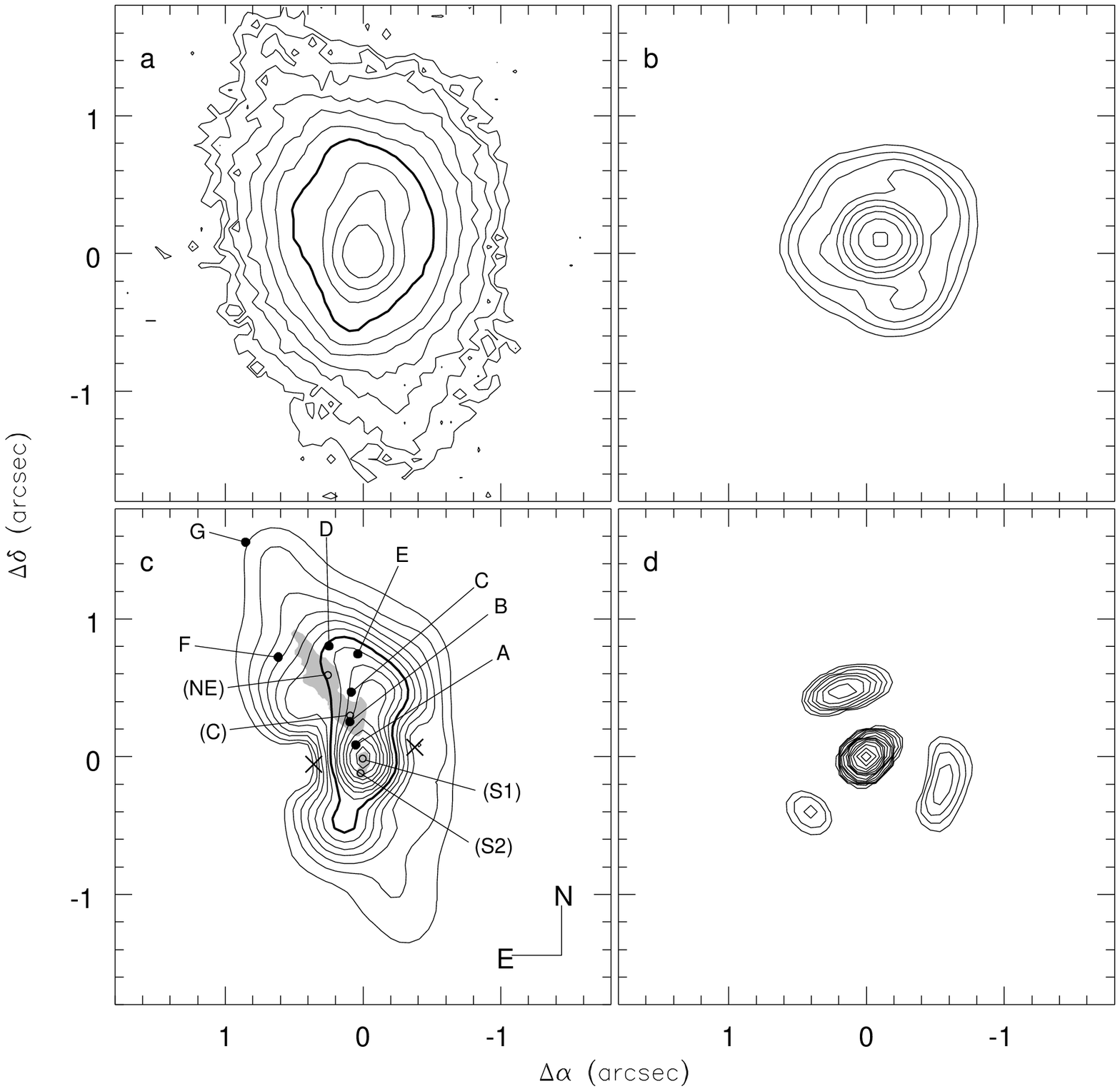}
\figcaption[f1.eps]{
  Contour maps of NGC 1068 at 11.7 $\mu$m. ($a$) The
  SAA image with contours starting at 0.63 Jy
  arcsec$^{-2}$, which is the 3 $\sigma$ level.
  The PSF reference image used for the Lucy-Richardson deconvolution is
  shown in ($b$).
  ($c$) The image of NGC 1068 after deconvolution.
  ($d$) The image of the test point source deconvolved with the
  same PSF reference.
  Five contours are plotted for a decade of surface brightness in a
  logarithmic scale.
  The thick contour is 10 Jy arcsec$^{-2}$ in ($a$) and ($c$).
  The filled circles in ($c$) show the locations and
  the names of the [\ion{O}{3}] emitting clouds \citep{Evans91}.
  Crosses are the locations of the M-band disklike structures
  \citep{Marco00}.
  Gray shade and open circles show the locations and
  the names of the 5 GHz continuum emission \citep{Gallimore96}.
\label{fig:R.contour}}
\end{figure}

\begin{figure}
\plotone{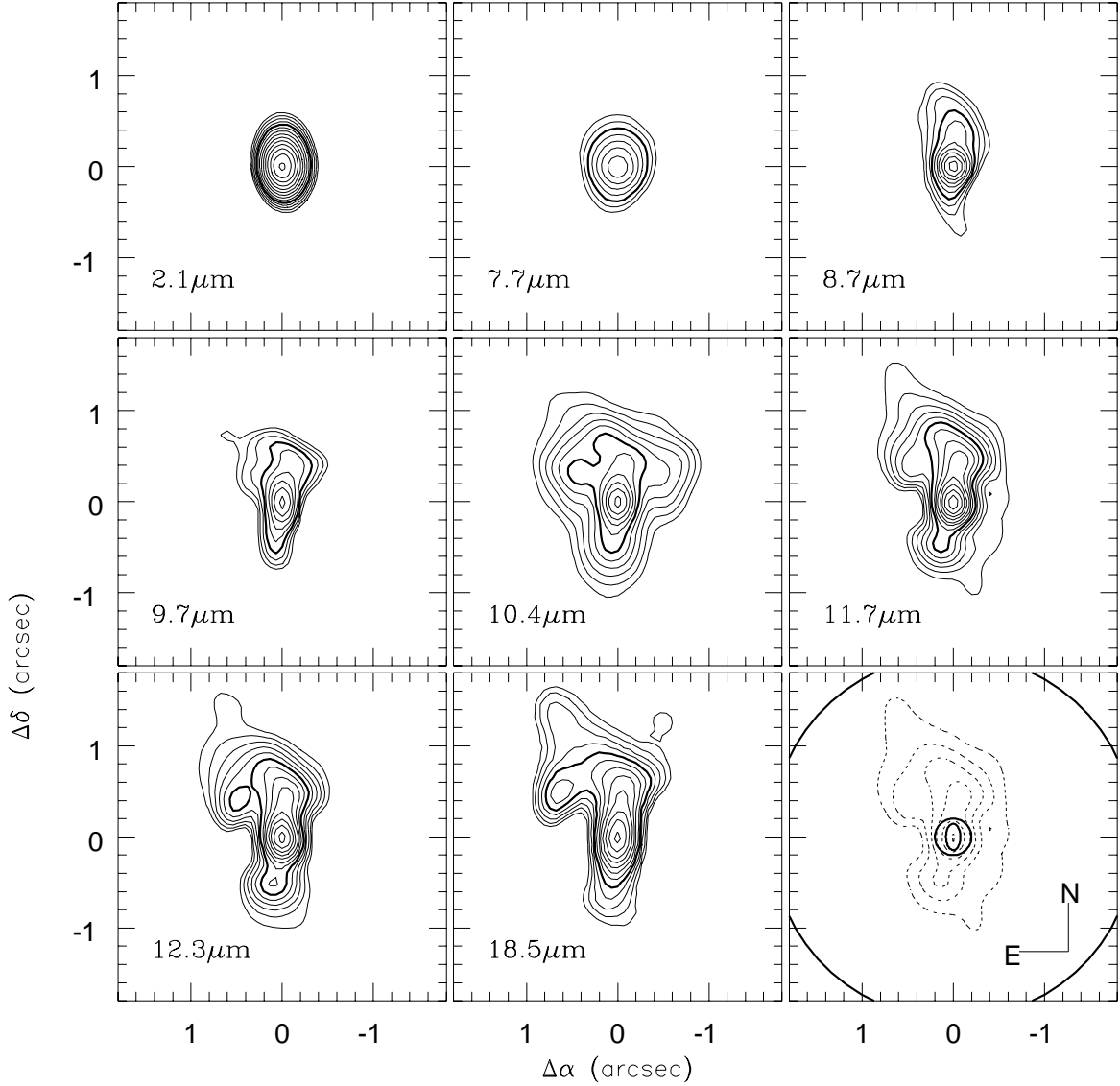}
\figcaption[f2.eps]{
  Deconvolved images of NGC 1068. Contours start above the 3 $\sigma$
  noise level. Five contours are spaced equally in a logarithmic scale
  in a decade. The 10 Jy arcsec$^{-2}$ (0.1 Jy arcsec$^{-2}$ for the
  2.1 $\mu$m image) contours are plotted with thick lines.
  The lower right-hand panel shows the apertures used for SED measurements
  overlaid on the 11.7 $\mu$m image.
\label{fig:Others.map}}
\end{figure}

Figure \ref{fig:R.contour} shows the effect of deconvolution
for the 11.7 $\mu$m image, starting with the image resulting from the SAA
method in Figure \ref{fig:R.contour}$a$.
Using the PSF in Figure \ref{fig:R.contour}$b$, the Lucy-Richardson
deconvolution resulted
in the image shown in Figure \ref{fig:R.contour}$c$.
Figure \ref{fig:Others.map} shows the deconvolved images of NGC 1068
in other wavelengths.
The deconvolved image of the deconvolution test star is
shown in Figure \ref{fig:R.contour}$d$.
There are three pointlike structures around
the star at the distance where the first diffraction ring would be.
Different rotation angles of the instrument derotator may
have caused the PSF difference.
Thus, structures around the central peak of NGC 1068 at the distance
of the artifact may be contaminated.
As the images of NGC 1068 are taken with different
rotator angles, influence of the artifactual structure in
the deconvolved image should be less than that for the
deconvolution test star. 

The structures seen in the SAA image are more prominent in the
deconvolved images.
Except for the 2.1 and 7.7 $\mu$m images, spatial
resolution of the images is the same in Figure
\ref{fig:Others.map}.
At 2.1 $\mu$m, spatial resolution of the image is limited by the
seeing.
As for the 7.7 $\mu$m image, the cause of the difference is not clear.

\subsection{Registration of the Central Peak}
\label{sec:regist}
To compare the MIR structures with those in other wavelengths, images
from other observations are overlaid on the MIR image.
\citet{Braatz93} measured relative position of the peak of NGC 1068 at
12.4 $\mu$m and in optical continuum.
Between the $K$ and $I$ bands, \citet{Marco97} measured relative
position of the peaks.
The positions of the 12.4 $\mu$m peak and the $K$-band peak agrees within
the precision of the measurements.
From our simultaneous observation in the $K$ band and in the MIR,
the peak of NGC 1068 in the two bands is measured to be at the same
position with an accuracy of $\pm$0\arcsec.3.
The accuracy was limited by the gear backlash on the movable beam splitter.
This will be improved in future observations.
\citet{Braatz93} also found that the 12.4 $\mu$m peak is consistent
with the apex of the [\ion{O}{3}] ionization cone
\citep{Evans91}.
Hereafter, the MIR peak and the $K$-band peak are assumed to be at the
position of the AGN.

Using the measurement of the $K$-band peak position \citep{Marco97} and
the optical peak position in the [\ion{O}{3}] image
\citep{Evans91}, the [\ion{O}{3}] emitting clouds are
registered on Figure \ref{fig:R.contour}$c$.
It also illustrates the positions of the 5 GHz components
\citep{Gallimore96} with the S1 component, which are most likely at
the AGN, overlaid on the MIR peak.
Moreover, peak positions of the NIR disklike structures \citep{Marco00}
are shown in the figure.

\section{Results}
\label{sec:Results}

\subsection{The Central Peak}
\label{sec:peak}
To investigate the spatial distribution of dust around the AGN, the spatial
extent of the central peak is measured and flux from the central peak
is fitted by a model of a modified graybody absorbed by the silicate
feature.
As described in \S \ref{sec:regist}, the K-band peak and the MIR peak
coincides with each other with the accuracy of $\pm$0.3 arcsec.
Moreover, the infrared peak is at the position where the AGN and the
dust torus reside.

\subsubsection{Flux Measurement}
An elliptical Gaussian is fitted on the central peak in the deconvolved
images. 
To avoid influence from the structures around the central peak, only
pixels with flux more than half of the peak are examined.
Results are shown in Table \ref{tab:FWHM}.
Except for the 2.1 and the 7.7 $\mu$m image, the
FWHM of the minor axis is about 0\arcsec.2.
As noted in \S \ref{sec:deconv}, deviation of the FWHM in the 2.1
and 7.7 $\mu$m image is not of the object.
The FWHM in the MIR of 0\arcsec.2 is slightly larger than the FWHM of
0\arcsec.13 for the deconvolution test star.
At all wavelengths, the central peak is elongated north and south.
Hereafter, we refer the central MIR emission region of
0\arcsec.29 $\times$ 0\arcsec.18 as the central elliptical region.
This is the FWHM fitted on the 11.7 $\mu$m image, which is made from
the largest number of frames.
At the distance of NGC 1068, the linear size of the central elliptical
region is 21 $\times$ 13 pc.

\begin{figure}
\plotone{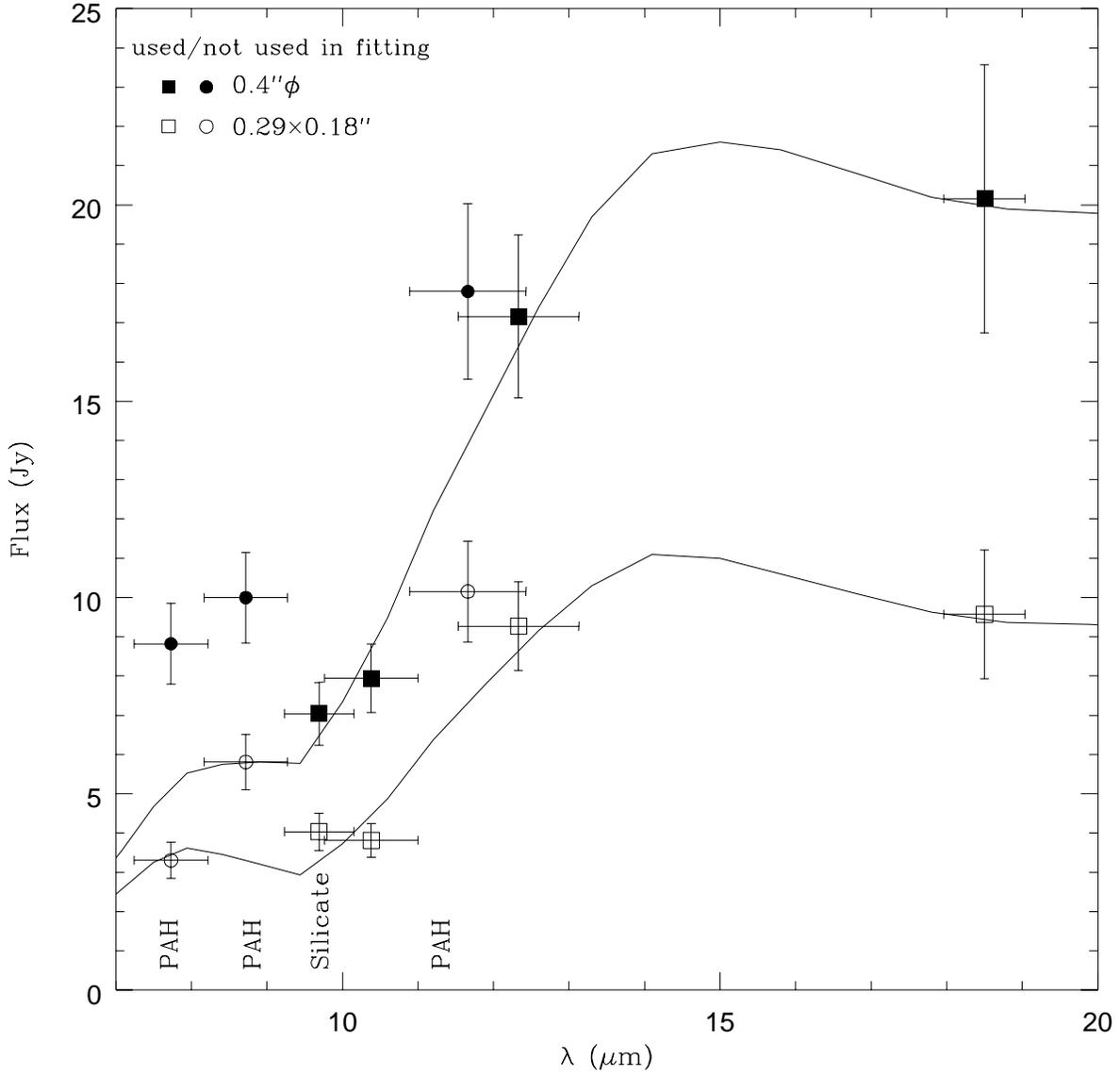}
\figcaption[f3.eps]{
  Flux in the apertures on the deconvolved images. The filled symbols show
  flux within the 0\arcsec.4 aperture. The open symbols show flux within
  the central elliptical region of 0\arcsec.29 $\times$ 0\arcsec.18.
  The upper curve shows the best-fit model for the 0\arcsec.4 aperture.
  The lower curve shows the best-fit model for the 0.\arcsec29 $\times$
  0.\arcsec18 aperture.
  Flux at 7.7, 8.7, and 11.7 $\mu$m,
  which is shown with circles, is not used in fitting owing to the
  possibility of PAH emission. Best-fit parameters are shown in Table
  \ref{tab:comps}.
  Wavelengths of peaks of the silicate features and the PAH features
  are also shown.
\label{fig:SEDbbfit}}
\end{figure}

Flux from the central peak is measured by applying apertures of three
sizes.
First, flux from the central elliptical region is measured.
The result is shown in the third column of Table \ref{tab:Flux}.
Note that the flux at 2.1 and 7.7 $\mu$m is lower limit because
of the spillover of the PSF in the deconvolved images.
Second, to avoid the spillover on the 7.7 $\mu$m image, we define a larger
aperture.
There is a secondary peak at 0\arcsec.4 north in the 11.7 $\mu$m image.
To exclude emission from the secondary peak, the diameter of the second
aperture is set to 0\arcsec.4.
The result is in the fourth column of Table \ref{tab:Flux}.
Figure \ref{fig:SEDbbfit} shows the measured spectral energy
distributions (SEDs) in the two
apertures.
The silicate absorption feature at 9.7 $\mu$m is seen.
Moreover, the 11.7 $\mu$m flux is larger than the 12.3 $\mu$m flux.
This suggests existence of the PAH emission feature at 11.3 $\mu$m.
Finally, for comparison with other observations, flux is also measured in
an aperture of 4\arcsec\ diameter. The result is in the fifth
column of Table \ref{tab:Flux}. This can be compared with the MIR
photometric observation of \citet{RiekeLow75} who used a diaphragm of
4\arcsec\ diameter. Our measurement in the 4\arcsec\ aperture is about
25\%--35\% brighter in the 7--13 $\mu$m bands than their results.
This is in reasonably good agreement considering uncertainty in the
measurements and variability of the target reported in the literature.
On the other hand, our measurement is about 30\% less bright than their
measurement around 18 $\mu$m. At the wavelength, the 4\arcsec\ 
diaphragm is comparable with the diffraction limit for the observations of
\citet{RiekeLow75}. This implies that there is extended emission
at around 18 $\mu$m.

\subsubsection{SED Fitting}
As shown by \citet{Roche91}, the absorption feature of silicate is
seen in spectra of AGNs. On the other hand, PAH emission features are
usually detected toward \ion{H}{2} region galaxies.
However, \citet{Voit92} pointed out possibility of PAH emission from
an AGN if the PAH carriers are sheltered from the intense radiation.
Here, the silicate absorption feature is measured with possibility of
the PAH emission features taken into account.

To measure the optical depth of the silicate feature, we conduct
least-squares
fitting of a modified graybody radiation absorbed by the silicate feature.
Integrating fitted modified graybody radiation,
intrinsic luminosity of the infrared emitting media can also be estimated.
Namely, SEDs in the MIR are fitted by model flux of
\begin{equation}
  I(\lambda) =
  \epsilon_{\lambda_0} \left( \frac{\lambda}{\lambda_0} \right)^{\alpha}
  \times B(T_{BB}, \lambda) \times e^{-\tau_{sil}(\lambda)}
\label{equ:bbmodel}\end{equation}
where $I(\lambda)$ is the model surface brightness at wavelength
$\lambda$.
The black body radiation $B(T_{BB}, \lambda)$ of temperature $T_{BB}$
is assumed to have the emissivity and beam filling factor product,
which is assumed to be $\epsilon_{\lambda_0}$ at $\lambda_0$ and
proportional to $\lambda^{\alpha}$.
The modified graybody radiation is absorbed by the silicate feature
of $\tau_{sil}(\lambda)$.
Fitting is done on the three free parameters: $T_{BB}$,
$\epsilon_{\lambda_0}$ at $\lambda_0$ = 10 $\mu$m, and
$\tau_{9.7 \mu m}$ = $\tau_{sil}(\lambda$ = 9.7 $\mu$m$)$.
Limited by the number of data points, $\alpha$ cannot be a free
parameter and is fixed at $-$1.
For the silicate feature, we assume the ``astronomical silicate'' by
\citet{LaorDraine93}\footnote{
  Optical property available on-line at
  \url{http://www.astro.princeton.edu/\~{}draine/.}
} with 0.1 $\mu$m radius. The tabulated absorption coefficient is
essentially the same for dust grains with a radius smaller than 0.5 $\mu$m
when normalized at 9.7 $\mu$m.
\citet{Mathis90} measured the interstellar extinction law.
The result is similar to the absorption coefficient of the
astronomical silicate in the MIR wavelength, but with fewer data
points.
As there is a possibility of contribution from PAH emission in the SEDs,
we excluded the 7.7, 8.6, and 11.3 $\mu$m images from fitting.
This also avoids influence of the PSF difference of the 7.7 $\mu$m image.

\begin{figure}
\plotone{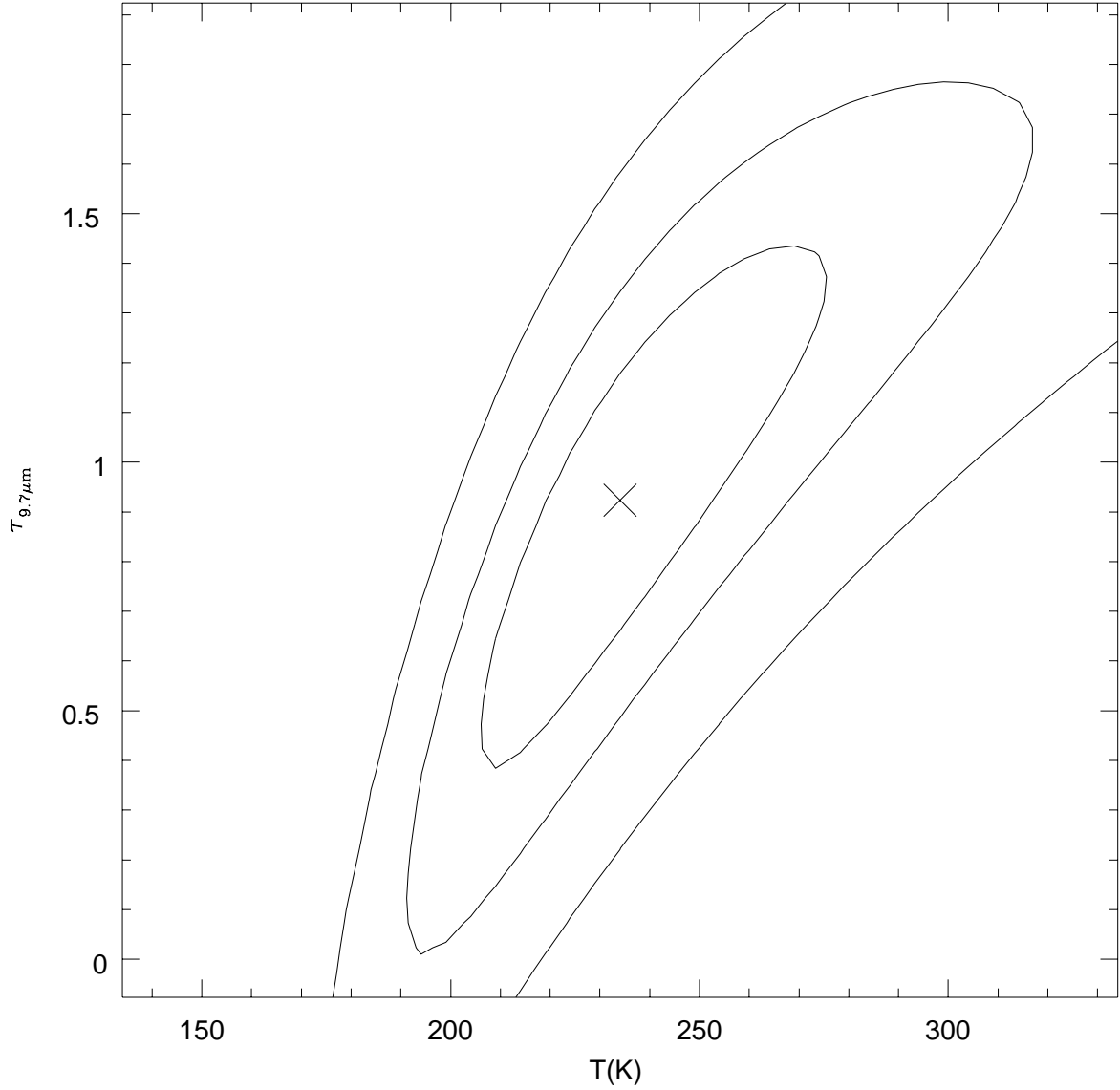}
\figcaption[f4.eps]{
  Confidence map of least-squares fitting on the SED for the
  0\arcsec.29 $\times$ 0\arcsec.18 aperture. Contours are for
  confidence levels of 68.3 \% (1 $\sigma$), 90 \%, and 99 \%.
\label{fig:chi2map}}
\end{figure}

The results of least-squares fittings are shown in Table
\ref{tab:comps}.
Figure \ref{fig:SEDbbfit} shows the model SEDs for the central
elliptical region and for the 0\arcsec.4 aperture.
For the larger aperture, the temperature is lower and the silicate
absorption is shallower.
The confidence map in Figure \ref{fig:chi2map} shows that the background
temperature and the optical depth of the silicate feature are related
although, for the 1 $\sigma$ level, the silicate absorption exists on
the SED.
The results do not change much with change of the assumed power law $\alpha$
of emissivity and filling factor product. When $\alpha$ is assumed as 0 or
$-$2, the optical depth $\tau_{9.7 \mu m}$ becomes larger or smaller,
respectively, about half of its 1 $\sigma$ error bar, compared with the
nominal value for $\alpha$ = 1.
Integrated graybody luminosity is 20 \% larger for $\alpha$ = 0
than for $\alpha$ = $-$1.
On the other hand, with $\alpha$ = $-$2, integrated luminosity
becomes 15 \% smaller than that for $\alpha$ = $-$1.

\citet{Roche84} observed the silicate absorption feature to be
$\tau_{9.7 \mu m}$ = 0.51 with the 4\arcsec.7 aperture.
This is consistent with our result for the 0\arcsec.4 aperture but
higher than that for the 4\arcsec\ aperture.
As a consequence of deconvolution, which gathers the light into the
central peak, the apertures on the deconvolved image are equivalent
to larger apertures on a direct image.

\subsection{PAH emission from the Central Peak}
As has been pointed out, for example, by \citet{Smith89}, the PAH features are
associated with starburst activity. In case they are detected from the
central peak, it would suggest that the PAH carriers near the AGN are
sheltered from the intense radiation as proposed by \citet{Voit92} and
that starburst activity is occurring near the AGN.

As Figure \ref{fig:SEDbbfit} shows, the observed flux
is underestimated by the model at 7.7, 8.7,
and 11.7 $\mu$m, wavelengths we excluded from the fitting.
Here, the excess emission is assumed to be the PAH emission features,
which peak at 7.7, 8.6, and 11.3 $\mu$m. For the
0\arcsec.29 $\times$ 0\arcsec.18 arcsec elliptical aperture, the 7.7
$\mu$m flux is not
underestimated, because the PSF is wider than the aperture.
Excess flux in the 0\arcsec.4 $\phi$ aperture is
  3.5 $\pm$ 1.0, 3.5 $\pm$ 1.0, and 2.6 $\pm$ 1.5 $\times$ 10$^{35}$ W for
  the 7.7, 8.6, and 11.3 $\mu$m feature
respectively. The contrast of the excess emission to the model continuum
is higher in smaller aperture.
This suggests that the PAH emission is from the central peak.

However, our result is inconsistent with other observations. Excess
flux in our 4\arcsec\ aperture is L$_{11.3\mu m}$ $\sim$
2 $\times$ 10$^{35}$ W. This is, for example, 7 times larger than
3.0 $\times$ 10$^{34}$ W that is measured by \citet{Sturm00} from the
spectrum taken by the Short Wavelength Spectrometer on board {\em
Infrared Space Observatory}. The single temperature continuum model adopted in
this work may be too simple, and uncertainty in relative calibration
between the filters may be too large to measure the excess luminosity.
A spatially resolved MIR spectrum is needed to clarify whether the PAH
features are emitted from the vicinity of the AGN.

\subsection{Emission Structures around the Central Peak}
\label{sec:structures}
The MIR deconvolved images show
structures elongated toward two directions. In
higher surface brightness, Figure \ref{fig:R.contour}$c$ shows an
elongation toward a P.A. of $-$10\arcdeg\ from the central peak.
The direction is consistent with the elongation imaged in the MIR by
\citet{Bock00}.
The northern part of the
elongation traces the western side of the ionization cone observed by
[\ion{O}{3}] emission \citep{Evans91}.
In lower surface brightness, another structure extends wider.
In the northern part, it elongates to a P.A. of 25\arcdeg.
This is consistent with the MIR images with wider FOV by
\citet{Braatz93} and \citet{Alloin00}.
The direction of the northeastern elongation coincides with that of
the jetlike structure observed by radio continua by, for example,
Gallimore et al. (1996).
Spatially resolved MIR spectrum may clarify the relations of the
emission structures.

\subsection{The Disklike Structure}
In L and M bands, \citet{Marco00} imaged disklike
structures in east and west of the central peak.
The locations are indicated by crosses in Figure \ref{fig:R.contour}$c$.
The MIR emission is constricted at the locations, and
essentially no MIR emission is detected.

\begin{figure}
\plotone{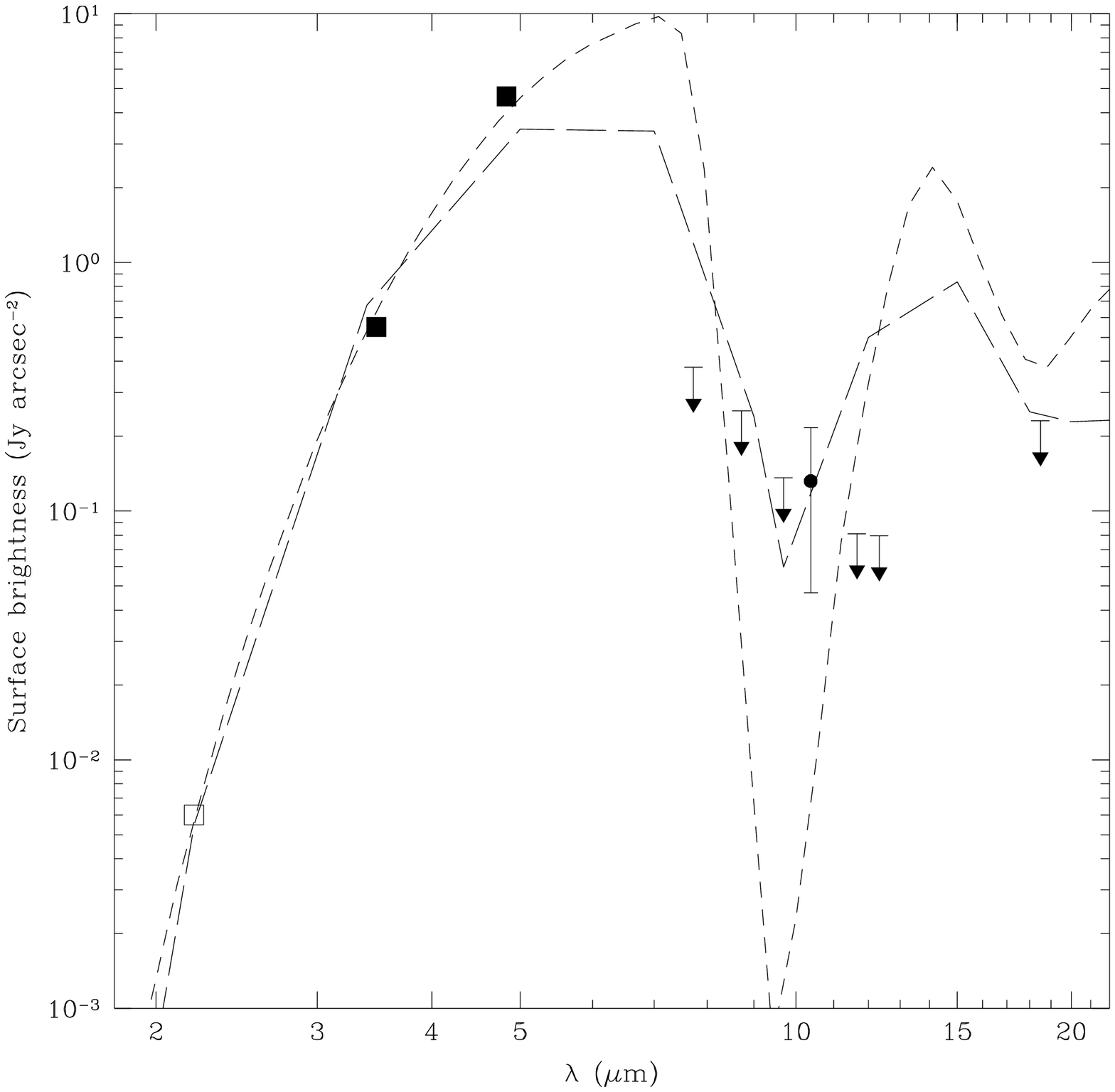}
\figcaption[f5.eps]{
  Average SED of the disklike structures in the east and the west of the
  central peak imaged by \citet{Marco00}. The filled squares are taken
  from \citet{Marco00}. The $K$-band brightness taken from \citet{Rouan98}
  is shown with an open square. No emission is detected in the MIR
  images. Upper limits of the 1 $\sigma$ level are shown. Curves show
  the models of modified graybody emission absorbed by the silicate
  feature. The short-dashed curve shows the model with 390 K background
  emission absorbed by astronomical silicate \citep{LaorDraine93} with
  $\tau_{9.7 \mu m}$ = 10. The long-dashed curve shows the model with
  1360 K background emission and the extinction law by \citet{Mathis90}
  with $\tau_{9.7 \mu m}$ = 5.
\label{fig:SEDdisk}}
\end{figure}

To assess the nature of the disklike structures, SED of the disklike
structures is plotted in Figure \ref{fig:SEDdisk} with help from the NIR
surface brightness taken from \citet{Marco00} and \citet{Rouan98}.
It is bright in the NIR but not in the MIR.
The SED is compared with the model of the modified graybody emission
absorbed by the silicate feature as presented by Equation
(\ref{equ:bbmodel}).
To attenuate the MIR emission, deep absorption of the silicate feature
is needed.
The absorption coefficient of the astronomical silicate and the interstellar
extinction law by \citet{Mathis90} differs in the NIR.
Therefore, we try different emission models adopting each optical
parameter.
When the astronomical silicate is assumed,
$\tau_{9.7 \mu m}$ has to be at least 10.
This corresponds to a hydrogen column density $N${\scriptsize H} at
least 1.6 $\times$ 10$^{23}$ cm$^{-2}$ assuming the Galactic extinction law
\citep{RocheAitken84, RocheAitken85, Predehl95}.
On the other hand, the model assuming the extinction law of
\citet{Mathis90} fits reasonably with $\tau_{9.7 \mu m}$ $\sim$ 5, or
$N${\scriptsize H} $\sim$ 8 $\times$ 10$^{22}$ cm$^{-2}$.
In the latter case $K$-band extinction is higher than the
astronomical silicate so that the $K$-band
brightness cannot be reproduced with $\tau_{9.7 \mu m}$ higher than 5.
In either case, the silicate absorption feature is too deep compared
with the absorption for the central peak.
Consequently, the disklike structures cannot be explained as
emission sources absorbed by dust.
The nature of the disklike structures remains unclear.

\section{Discussion}
\label{sec:Discussion}
\subsection{MIR Radiation from the Central Peak}
On the 11.7 $\mu$m image, the central peak has an FWHM of
0\arcsec.29 $\times$ 0\arcsec.18, or a linear scale of 21 $\times$ 13
pc, elongated
north and south.
From the direction of the [\ion{O}{3}] cone \citep{Evans91},
the axis of the torus is also in the north-south direction.
It implies that the plausible torus has a diameter comparable or
smaller than 13 pc.

Fitting on the measurement of \citet{RiekeLow75}, \citet{PierKrolik93}
modeled emission from the AGN obscuring torus.
As shown in their figure, NGC 1068 emits a considerable portion of its
energy in the MIR.
Thus, intrinsic luminosity of the torus can be estimated integrating
the fitted modified graybody radiation.
Integrating the modified graybody emission over the wavelength and
surface of the ellipsoid, intrinsic luminosity of the modified graybody
emission is calculated to be 3.3 $\times$ 10$^{37}$ W.
This is comparable to the intrinsic X-ray emission luminosity
estimated by \citet{Iwasawa97}.

\subsection{Dust Absorption toward the Central Source}
The measured silicate absorption is compared with observations in other
wavelengths.
It is known that the extinction law toward the Galactic center is
different from the one for the local inter stellar matter (ISM).
The measured optical depth for the central elliptical region of
$\tau_{9.7 \mu m}$ = 0.9 $\pm$ 0.3 corresponds to an $A_V$ $\sim$ 8 mag
with the extinction law for the local ISM \citep{RocheAitken84} or
17 $\pm$ 6 mag with that toward the Galactic center
\citep{RocheAitken85}.
The corresponding hydrogen column density is calculated as
$N${\scriptsize H} $\sim$ 1.4 or 3.0 $\times$ 10$^{22}$ cm$^{-2}$
\citep{Predehl95}.

As for the 3.4 $\mu$m feature of carbonaceous dust,
\citet{Pendleton94} measured the extinction law $A_V/\tau_{3.4}$
toward the Galactic center and for the local ISM.
The values are again different by a factor of 2
although, optical depth ratio $\tau_{3.4 \mu m}/\tau_{9.7 \mu m}$
calculated from the extinction ratios only differs $\sim$ 20\%.
Namely, it is $\sim$ 0.06 toward the Galactic
center and $\sim$ 0.074 for the local ISM.
Using the mean value, $\tau_{9.7 \mu m}$ $\sim$ 0.9
corresponds to $\tau_{3.4\mu m}$ $\sim$ 0.06.
This is half of the observed optical depth \citep{Imanishi97}.
As has been suggested by \cite{Imanishi00}, the discrepancy can be
attributed to a temperature structure of the absorbing dust along the line
of sight.
Comparing models, the wavelength dependency of the dust features may
constrain the temperature structure of dust.
It should also be noted that there is evidence that the extinction law
toward Seyfert nuclei is different from the one in our Galaxy
throughout all wavelengths observable from the ground \citep{Pitman00,
Crenshaw01, Maiolino01a, Imanishi01}.
There is a good chance of the ratio $\tau_{3.4 \mu m}/\tau_{9.7 \mu
m}$ being different from our Galaxy, because the difference of the
extinction is attributed to different dust size distribution
\citep{Maiolino01b, Imanishi01}.
The two possibilities, the temperature gradient and the different
optical depth ratio, are not distinguishable from this work.
Correlation of $\tau_{3.4 \mu m}/\tau_{9.7 \mu m}$ has to be
investigated toward different classes of objects.

\section{Conclusion}
\label{sec:Conclusion}
We observed the central region of the Seyfert 2 galaxy NGC 1068 using
the MIRTOS on the Subaru Telescope.
The oversampling pixel scale accompanied by the SAA method enables us
to obtain high spatial resolution images of about 0\arcsec.13 FWHM after
Lucy-Richardson deconvolution in the MIR.
The central peak of NGC 1068 in the MIR is 0\arcsec.3 FWHM north-south
and 0\arcsec.2 FWHM east-west.
It corresponds to a linear scale of 21 $\times$ 13 pc.
With its axis has been observed to be in the north-south direction,
the plausible torus has a diameter comparable or less than 13 pc.
Overall structure elongates to a P.A. of $-$10\arcdeg\ with another
elongation of lower surface brightness at a P.A. of 20\arcdeg.
The six filters in the 10 $\mu$m band along with the 18.5 $\mu$m
filter produced the detailed SED of the central peak.
The SED of the central elliptical region of 0\arcsec.3 $\times$ 0\arcsec.2 is
fitted to have a modified graybody emission of 234 K with silicate
absorption feature $\tau_{9.7 \mu m}$ = 0.9 $\pm$ 0.3.
Intrinsic luminosity of the central peak is about 3 $\times$ 10$^{37}$ W.
The MIR silicate absorption feature is half of that estimated from the
NIR spectroscopy \citep{Imanishi97} when the Galactic extinction law
is assumed.
This can be explained by a temperature gradient of dust along the
line of sight as \citet{Imanishi00} suggested, or by different dust-size
distribution as Maiolino et al. (2001b) and \citet{Imanishi01}
pointed out.
From the central source, there is excess emission at the wavelengths
of the PAH emission features.
Uncertainty of the measurement is huge, and we have to wait
for a high spatial resolution spectrum.
However, this suggests that the PAH carrier is shielded from high
energy radiation from the AGN \citep{Voit92} and the existence of
the nuclear starburst.
No MIR emission is detected as counterparts for the NIR disklike
structures \citep{Alloin00}.
The nature of the structure is still unclear.

\acknowledgments
We are grateful to Danielle Alloin for valuable suggestions during
her stay at National Astronomical Observatory of Japan in Tokyo,
Japan.
We thank Naoto Kobayashi and Masayuki Akiyama for many
constructive discussions.
Also, we thank Masatoshi Imanishi and the anonymous referee who
suggested us many viewpoints to investigate our data.
The staff at the Subaru Telescope enabled to realize the observations
with the newly constructed instrument.
We appreciate their enormous efforts.
D. T. has been supported by the Research Fellowship for Young Scientists
from the Japan Society for the Promotion of Science.

\clearpage

\clearpage

\begin{deluxetable}{lcr@{ $\pm$ }lr@{ $\pm$ }lr@{ $\pm$ }l}
\tablecaption{
  Filter Path Bands and Observed Flux of NGC 1068 in Apertures on
  Images Deconvolved by the Lucy-Richardson Algorithm
  \label{tab:Flux}}
\tablewidth{0pc}
\tablehead{
  \colhead{$\lambda$} & \colhead{$\Delta\lambda$} &
  \multicolumn{2}{c}{0\arcsec.29 $\times$ 0\arcsec.18\tablenotemark{a}} &
  \multicolumn{2}{c}{0\arcsec.4 $\phi$} & \multicolumn{2}{c}{4\arcsec $\phi$} \\
  \colhead{($\mu$m)} & \colhead{($\mu$m)} &
  \multicolumn{2}{c}{(Jy)} & \multicolumn{2}{c}{(Jy)} & \multicolumn{2}{c}{(Jy)}
}
\startdata
      2.12 & 0.36 & 0.218 & 0.027\tablenotemark{b}
                  &  0.432 & 0.054\tablenotemark{b} &  0.560 & 0.070 \\ 
      7.73 & 0.70 &  3.31 & 0.46\tablenotemark{b}
                  &  8.8   & 1.0   & 16.2   & 1.9   \\ 
      8.72 & 0.78 &  5.81 & 0.70  & 10.0   & 1.2   & 16.5   & 1.9   \\ 
      9.69 & 0.94 &  4.03 & 0.48  &  7.04  & 0.80  & 16.1   & 1.8   \\ 
     10.38 & 1.02 &  3.82 & 0.43  &  7.94  & 0.87  & 22.8   & 2.5   \\ 
     11.66 & 1.16 & 10.2  & 1.3   & 17.8   & 2.2   & 33.9   & 4.3   \\ 
     12.33 & 1.18 &  9.3  & 1.1   & 17.2   & 2.1   & 33.3   & 4.0   \\ 
     18.50 & 1.08 &  9.6  & 1.6   & 20.2   & 3.4   & 46.6   & 7.9   \\ 
\enddata
\tablecomments{Error bars include uncertainty in the conversion factors.}
\tablenotetext{a}{Ellipse with P.A. of 0\arcdeg.}
\tablenotetext{b}{The values are lower limits because the PSF is
  larger than the aperture.}
\end{deluxetable}

\begin{deluxetable}{lcccccccc}
\tablecaption{
  Number of Observed Nodding Sets for NGC 1068
  \label{tab:obslog:NGC}}
\tablewidth{0pc}
\tablehead{
  \colhead{UT} &
  \colhead{2.1 $\mu$m} &
  \colhead{7.7 $\mu$m} &
  \colhead{8.7 $\mu$m} &
  \colhead{9.7 $\mu$m} &
  \colhead{10.4 $\mu$m} &
  \colhead{11.7 $\mu$m} &
  \colhead{12.3 $\mu$m} &
  \colhead{18.5 $\mu$m}
}
\startdata
Dec 31
  & \nodata 
  & \nodata
  & \nodata
  & \nodata
  & 4
  & 8
  & \nodata
  & \nodata \\
Jan 9
  & 24
  & 2
  & 2
  & 2
  & 4
  & 4
  & 6
  & 4 \\
Jan 18
  & 22
  & \nodata
  & \nodata
  & \nodata
  & \nodata
  & 8
  & 12
  & 12 \\
\enddata
  \tablecomments{One nodding set corresponds to 3 s of integration
    time on-source.}
\end{deluxetable}

\begin{deluxetable}{lcccccccc}
\tablecaption
  {Number of Observed Nodding Sets for PSF Reference Stars
  \label{tab:obslog:psfref}}
\tablewidth{0pc}
\tablehead{
  \colhead{UT} &
  \colhead{2.1 $\mu$m} &
  \colhead{7.7 $\mu$m} &
  \colhead{8.7 $\mu$m} &
  \colhead{9.7 $\mu$m} &
  \colhead{10.4 $\mu$m} &
  \colhead{11.7 $\mu$m} &
  \colhead{12.3 $\mu$m} &
  \colhead{18.5 $\mu$m}
}
\startdata
Dec 31
  & \nodata
  & \nodata
  & \nodata
  & \nodata
  & 2\tablenotemark{a}
  & 2\tablenotemark{a}
  & \nodata
  & \nodata \\
Jan 9
  & 2\tablenotemark{b}
  & 2\tablenotemark{c}
  & 2\tablenotemark{c}
  & 2\tablenotemark{c}
  & 6\tablenotemark{c}
  & 2\tablenotemark{c}
  & 2\tablenotemark{c}
  & \nodata \\
Jan 18
  & 4\tablenotemark{d}
  & \nodata
  & \nodata
  & \nodata
  & \nodata
  & 2\tablenotemark{e}
  & 2\tablenotemark{e}
  & 4\tablenotemark{e} \\
\enddata
\tablenotetext{a}{$\alpha$ Ari.}
\tablenotetext{b}{HD 40335.}
\tablenotetext{c}{$\alpha$ CMa.}
\tablenotetext{d}{GJ 105.5.}
\tablenotetext{e}{$\beta$ And.}
\end{deluxetable}

\begin{deluxetable}{lccccc}
\tablecaption
  {Number of Observed Nodding Sets for Deconvolution Test Stars
  \label{tab:obslog:dectst}}
\tablewidth{0pc}
\tablehead{
  \colhead{UT} &
  \colhead{2.1 $\mu$m} &
  \colhead{10.4 $\mu$m} &
  \colhead{11.7 $\mu$m} &
  \colhead{18.5 $\mu$m}
}
\startdata
Dec 31
  & \nodata
  & 2\tablenotemark{a}
  & 2\tablenotemark{a}
  & 4\tablenotemark{a} \\
Jan 9
  & 2\tablenotemark{b}
  & \nodata
  & 5\tablenotemark{a}
  & \nodata \\
\enddata
\tablenotetext{a}{$\beta$ And.}
\tablenotetext{b}{GJ 105.5.}
\end{deluxetable}

\begin{deluxetable}{lcccccccc}
\tablecaption
  {Number of Observed Nodding Sets for Flux Reference Stars
  \label{tab:obslog:lumref}}
\tablewidth{0pc}
\tablehead{
  \colhead{UT} &
  \colhead{2.1 $\mu$m} &
  \colhead{7.7 $\mu$m} &
  \colhead{8.7 $\mu$m} &
  \colhead{9.7 $\mu$m} &
  \colhead{10.4 $\mu$m} &
  \colhead{11.7 $\mu$m} &
  \colhead{12.3 $\mu$m} &
  \colhead{18.5 $\mu$m}
}
\startdata
Dec 31
  & \nodata
  & \nodata
  & \nodata
  & \nodata
  & 4\tablenotemark{a}
  + 2\tablenotemark{b}
  & 28\tablenotemark{a}
  + 26\tablenotemark{b}
  & \nodata
  & \nodata \\
Jan 10
  & 54\tablenotemark{c}
  + 12\tablenotemark{d}
  & 2\tablenotemark{b}
  + 2\tablenotemark{e}
  & 2\tablenotemark{b}
  + 8\tablenotemark{e}
  & 2\tablenotemark{b}
  + 4\tablenotemark{e}
  & 2\tablenotemark{b}
  + 14\tablenotemark{e}
  & 28\tablenotemark{b}
  + 14\tablenotemark{e}
  & 2\tablenotemark{b}
  + 2\tablenotemark{e}
  & 4\tablenotemark{b} \\
Jan 19
  & 6\tablenotemark{c}
  & \nodata
  & \nodata
  & \nodata
  & \nodata
  & 12\tablenotemark{b}
  & 2\tablenotemark{b}
  & 6\tablenotemark{b} \\
\enddata
\tablenotetext{a}{$\alpha$ Ari.}
\tablenotetext{b}{$\beta$ And.}
\tablenotetext{c}{GJ 105.5.}
\tablenotetext{d}{HD 40335.}
\tablenotetext{e}{$\alpha$ CMa.}
\end{deluxetable}

\begin{deluxetable}{lcccc}
\tablecaption{FWHM of the Central Peak after Lucy-Richardson Deconvolution
  \label{tab:FWHM}}
\tablewidth{0pc}
\tablehead{
  \colhead{$\lambda$} &
  \colhead{Major axis} &
  \colhead{Minor axis} &
  \colhead{P.A. of major axis} &
  \colhead{Reference\tablenotemark{a}} \\
  \colhead{($\mu$m)} &
  \colhead{(arcsec)} &
  \colhead{(arcsec)} &
  \colhead{(degrees)} &
  \colhead{(arcsec)}
}
\startdata
    2.1 & 0.74 & 0.54 & $+$0.6 & 0.48 \\ 
    7.7 & 0.36 & 0.33 & $+$0.0 & \nodata  \\ 
    8.7 & 0.26 & 0.21 & $+$1.2 & \nodata  \\ 
    9.7 & 0.39 & 0.21 & $+$0.1 & \nodata  \\ 
   10.4 & 0.35 & 0.23 & $-$0.6 & 0.11 \\ 
   11.7 & 0.29 & 0.18 & $+$0.6 & 0.13 \\ 
   12.3 & 0.30 & 0.19 & $+$0.2 & \nodata  \\ 
   18.5 & 0.43 & 0.22 & $+$0.2 & \nodata  \\ 
\enddata
\tablenotetext{a}{FWHM of the deconvolution test star after
  Lucy-Richardson deconvolution.}
\end{deluxetable}

\begin{deluxetable}{lr@{ $\pm$ }lr@{ $\pm$ }lr@{ $\pm$ }lc}
\tablecaption{Best-Fit Parameters on Central Peak
\label{tab:comps}}
\tablewidth{0pc}
\tablehead{
  \colhead{Aperture} &
  \multicolumn{2}{c}{$T_{BB}$} &
  \multicolumn{2}{c}{} &
  \multicolumn{2}{c}{} &
  \colhead{} \\
  \colhead{arcsec} &
  \multicolumn{2}{c}{K} &
  \multicolumn{2}{c}{$\tau_{9.7 \mu m}$} &
  \multicolumn{2}{c}{$\log_{10}(\epsilon_{10\mu m})$} &
  \colhead{$\chi^{2}$}
}
\startdata
  0.29 $\times$ 0.18 ellipse & 234&22 & 0.92&0.34 & -0.98&0.14 & 5.1 \\
  0.4$\phi$                & 219&19 & 0.72&0.33 & -1.06&0.14 & 1.5 \\
  4$\phi$                  & 201&15 & 0.12&0.30 & -2.64&0.14 & 0.4 \\
\enddata
\tablecomments{
  Blackbody radiation is assumed to have emissivity and filling factor
  product proportional to $\lambda^{-1}$.
  SEDs are fitted with the three free parameters on flux at
  9.7, 10.4, 12.3, and 18.5 $\mu$m.
}
\end{deluxetable}

\begin{thebibliography}{}

\bibitem[Alloin et al.(2000)]{Alloin00}
  Alloin, J., et al. 2000, \aap, 363, 926

\bibitem[Antonucci(1993)]{Antonucci93}
  Antonucci, R. 1993, \araa, 31, 473

\bibitem[Antonucci \& Miller(1985)]{Antonucci85}
  Antonucci, R. \& Miller, J. S. 1985, \apj, 297, 621

\bibitem[Bock et al.(2000)]{Bock00}
  Bock, J. J., et al. 2000, \aj, 120, 2904

\bibitem[Bottinelli et al.(1990)]{Bottinelli90}
  Bottinelli, L., et al. 1990, \aaps, 82, 391

\bibitem[Braatz et al.(1993)]{Braatz93}
  Braatz, J. A., et al. 1993, \apj, 409, L5

\bibitem[Cohen et al.(1995)]{Cohen95}
  Cohen, M., et al. 1995, \aj, 110, 275

\bibitem[Cornwell \& Evans(1985)]{CornwellEvans85}
  Cornwell, T. J. \& Evans, K. F. 1985, \aap, 143, 77

\bibitem[Crenshaw et al.(2001)]{Crenshaw01}
  Crenshaw, D. M., Kraemer, S. B., Bruhweiler, F. C., \& Ruiz, J. R.
  2001, \apj, 555, 633

\bibitem[Elias et al.(1982)]{Elias82}
  Elias, J. H., et al. 1982, \aj, 87, 1029

\bibitem[Evans et al.(1991)]{Evans91}
  Evans, I. N., et al. 1991, \apj, 369, L27

\bibitem[Gallimore, Baum, \& O'Dea(1996)]{Gallimore96}
  Gallimore, J. F., Baum, S. A., \& O'Dea, C. P. 1996, \apj, 464, 198

\bibitem[Imanishi(2000)]{Imanishi00}
  Imanishi, M. 2000, \mnras, 319, 331

\bibitem[Imanishi(2001)]{Imanishi01}
  Imanishi, M. 2001, \aj, 121, 1927

\bibitem[Imanishi et al.(1997)]{Imanishi97}
  Imanishi, M., et al. 1997, \pasj, 49, 69

\bibitem[Iwasawa et al.(1997)]{Iwasawa97}
  Iwasawa, K., et al. 1997, \mnras, 289, 443

\bibitem[Laor \& Draine(1993)]{LaorDraine93}
  Laor, A, \& Draine, B. T. 1993, \apj, 402, 441

\bibitem[Lucy(1974)]{Lucy74}
  Lucy, L. B. 1974, \aj, 79, 745

\bibitem[Maiolino, Marconi, \& Oliva(2001a)]{Maiolino01a}
  Maiolino, R., Marconi, A., \& Oliva, E. 2001a, \aap, 365, 28

\bibitem[Maiolino, Marconi, \& Oliva(2001b)]{Maiolino01b}
  Maiolino, R., Marconi, A., \& Oliva, E. 2001b, \aap, 365, 37

\bibitem[Marco \& Alloin(2000)]{Marco00}
  Marco, O. \& Alloin, D. 2000, \aap, 353, 465

\bibitem[Marco et al.(1997)]{Marco97}
  Marco, O., Alloin, D., \& Beuzit, J. L. 1997, \aap, 320, 399

\bibitem[Mathis(1990)]{Mathis90}
  Mathis, J. S. 1990, \araa, 28, 37

\bibitem[Pendleton et al.(1994)]{Pendleton94}
 Pendleton, Y. J., Sandford, S. A., Allamandola, L. J., Tielens, A. G.
 G. M., \& Sellgren, K. 1994, \apj, 437, 683

\bibitem[Pier \& Krolik(1993)]{PierKrolik93}
  Pier, E. A. \& Krolik, J. H. 1993, \apj, 418, 673

\bibitem[Pitman, Clayton, \& Gordon(2000)]{Pitman00}
  Pitman, K. M., Clayton, G. C., \& Gordon, K. D. 2000, \pasp, 112,
  537

\bibitem[Predehl \& Schmitt(1995)]{Predehl95}
  Predehl, P. \& Schmitt, J. H. M. M. 1995, \aap, 293, 889

\bibitem[Rieke et al.(1985)]{Rieke85}
  Rieke, G. H., et al. 1984, \aj, 90, 900

\bibitem[Rieke \& Low(1975)]{RiekeLow75}
  Rieke, G. H. \& Low, F. J. 1975, \apj, 199, L13

\bibitem[Roche \& Aitken(1984)]{RocheAitken84}
  Roche, P. F. \& Aitken, D. K. 1984, \mnras, 208, 481

\bibitem[Roche \& Aitken(1985)]{RocheAitken85}
  Roche, P. F. \& Aitken, D. K. 1985, \mnras, 215, 425

\bibitem[Roche et al.(1984)]{Roche84}
  Roche, P. F., Aitken, D. K., Phillips, M. M., \& Whitmore, B. 1984,
  \mnras, 207, 35

\bibitem[Roche et al.(1991)]{Roche91}
  Roche, P. F., Aitken, D. K., Smith, C. H., \& Martin, J. W. 1991,
  \mnras, 248, 606

\bibitem[Rouan et al.(1998)]{Rouan98}
  Rouan, D., et al. 1998, \aap, 339, 687

\bibitem[Smith, Aitken, \& Roche(1989)]{Smith89}
  Smith, C. H., Aitken, D. K., \& Roche, P. F. 1989, \mnras, 241, 425

\bibitem[Sturm et al.(2000)]{Sturm00}
  Sturm, E., et al. 2000, \aap, 358, 481

\bibitem[Tokunaga(1984)]{Tokunaga84}
  Tokunaga, A. T. 1984, \aj, 89, 172

\bibitem[Tomono(2000)]{Tomono00}
  Tomono, D. 2000, Ph. D. thesis, Univ. Tokyo

\bibitem[Tomono, Doi, \& Nishimura(2000)]{Tomonoetal00}
  Tomono, D., Doi, Y., \& Nishimura, T. 2000, Proc. SPIE, 4008, 853

\bibitem[Voit(1992)]{Voit92}
  Voit, G. M. 1992, \mnras, 258, 841

\end{thebibliography}
\end{document}